\documentclass[amsmath,amssymb,twocolumn,notitlepage]{revtex4-1} 
	\pdfoutput=1  
	\usepackage{graphicx}
    \usepackage{dcolumn}
    \usepackage{bm}
    \usepackage[version=3]{mhchem}
    \usepackage{color}

\begin{document}

\title{Compact High-Q Optical Comb based on a Photonic Harmonic Potential}

\author{Sylvain Combri\'{e}}
\author{Ga\"{e}lle~Lehoucq}
\author{Gregory Moille}
\email{current affiliation: National Institute of Standards and Technology, Gaithersbourg, MD, USA / University of Maryland, College Park, MD, USA.}
\author{Aude Martin}
\author{Alfredo De Rossi}
\email{alfredo.derossi@thalesgroup.com}
\affiliation{Thales Research and Technology France, 1 avenue Augustin Fresnel, 91120 Palaiseau, France} 

\date{\today}

\begin{abstract}
An effective harmonic potential for photons is achieved in a photonic crystal structure, owing to the balance of the background dispersion and a bichromatic potential. Consequently, ultra-compact resonators with several equi-spaced resonances and high loaded Q factors (0.7 million) are demonstrated. 
A detailed statistical analysis is carried out by exploiting the complex reflection spectra measured with Optical Coherent Tomography. The log-normal distribution of the intrinsic Q-factors peaks at 3 million. The device is made of $Ga_{0.5}In_{0.5}P$ in order to suppress the two photon absorption in the Telecom spectral range considered here. This is crucial to turn the strong localization of light into ultra-efficient parametric interactions.
\end{abstract}

\maketitle

\section{introduction}
Multi-mode optical microcavities have enabled the emergence of compact optical sources such as parametric oscillators~\cite{kippenberg_kerr-nonlinearity_2004} and nonlinear combs~\cite{matsko_review_2005} for quantum information~\cite{grassani_micrometer-scale_2015,preble_-chip_2015,reimer_generation_2016,li_efficient_2016} and metrology~\cite{kippenberg_microresonator-based_2011,huang_broadband_2016}.
Efficient nonlinear interactions take place, owing to a combination of large Q factor, small modal volume and large material index~\cite{kippenberg_kerr-nonlinearity_2004,ji_breaking_2016,pu_efficient_2016,xuan_high-q_2016}. In any case, the control of the frequency spacing of the resonances is critical.\\
As for the archetypal Fabry-Perot, the optical comb involves higher order modes in microdisks, microspheres, toroids and microrings, where the frequency spacing tends to an ideal comb and can be precisely tuned through dispersion engineering (see e.g. \cite{matsko_review_2005}). Therefore, scaling down the size of resonators is challenged by the increasing role of the transverse confinement in the dispersion. This happens well before the onset of diffraction losses. Let us consider the modes of an idealized Fabry-Perot resonator with length $L$ with frequencies: $\nu_n =c \sqrt{(\pi n/L)^2+k_t^2}$, where $c$ is the speed of light in the material. As the mode number $n$ decreases, the spacing will increasingly deviate from the equal spacing due to the transverse confinement $k_t$. This is a general result, which applies to any resonator based on waves circulating in a weakly dispersive medium.\\ 
The fundamental question which is addressed here is whether a comb-like mode structure could be created starting from the \textit{fundamental order mode}. Clearly, this requires a radically different concept for the confinement of light in a dielectric, where the dispersion is radically different from that of the homogeneous medium. This is the case of Photonic Crystals where it was pointed out\cite{john_strong_1987,notomi_theory_2000,painter_wannier-like_2003,savona_electromagnetic_2011} that photons near the band edge follow a free-electron like dispersion $\hbar\omega\approx \hbar\omega_0+\hbar^2k^2/2m^*$, where $m^*$ is the effective mass for electrons. A line of $L$ missing holes in a triangular lattice of holes drilled in a sub-wavelength dielectric slab creates an optical resonator described by a Fabry-Perot model with effective length $\approx(1+L)a$ and the dispersion given by the underlying waveguide\cite{lalanne_photon_2008,kim_characteristics_2004}. The localisation of photons in such a system is equivalent of that of electrons in a square potential well. Consequently, levels are spaced following a quadratic rule, e.g. $\hbar\omega_n\propto n^2$, as shown in Fig.~\ref{fig:CavityHarmonicPotential}a (left). In contrast, the creation of a parabolic potential for photons would create a system equivalent to the harmonic oscillator in quantum mechanics, with rigorously equi-spaced levels.\\
A parabolic modulation of the lattice parameters has been introduced in Photonic Crystals long ago\cite{barclay_nonlinear_2005}, aiming at the suppression of radiative losses~\cite{srinivasan_momentum_2002} for high-Q resonators~\cite{tanaka_design_2008,notomi_ultrahigh-q_2008}. This approach was so successful that Q factors exceeding $10^7$ have been achieved~\cite{asano_photonic_2017}. This design strategy is also particularly effective with structures made of low index materials, such as $SiN$\cite{mccutcheon_design_2008}, $SiO_2$\cite{gong_photonic_2010} or $III-V$ semiconductors embedded in a solid state cladding\cite{bazin_design_2014}. In the context of opto-mechanic crystals, a parabolic modulation has been used to tailor optical and mechanical modes and it has been noted that this introduces an effective quasi-harmonic potential\cite{eichenfield_optomechanical_2009}. Higher order modes in similar structures are briefly discussed in Ref.~\cite{quan_deterministic_2011}, but their frequency spacing has not been addressed.\\
Here we demonstrate an effective harmonic potential for photons using a bichromatic photonic crystal. The experimental evidence is given through the statistical analysis of the frequency spacing of high-Q resonances over 68 cavities.  For each structure, the complex reflection spectra is measured using Optical Coherent Tomography with a resolution of 20 MHz. Moreover, each of these resonators have many, typically four modes, with loaded Q factor in the $10^5$ to $7\times10^5$ range and intrinsic Q factor well above one million. Equi-spaced resonances in a very compact structure should lead to ultra-strong nonlinear interactions, particularly resonantly enhanced Four Wave Mixing and parametric oscillation.\\
\section{Photonic harmonic potential}
%
%
\begin{figure}[htbp]
	\centering
	\includegraphics[width=\linewidth]{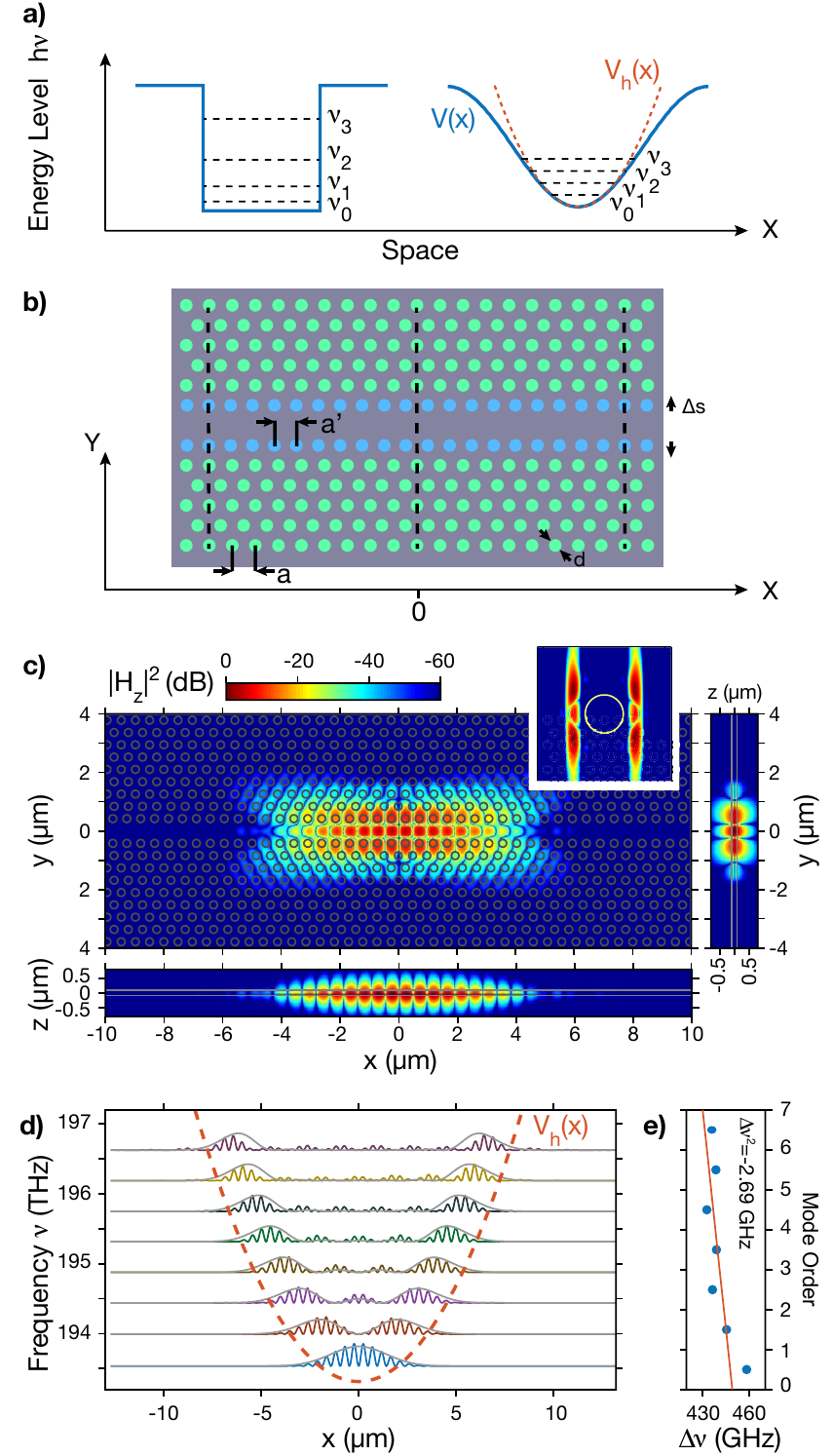}
	\caption{$a)$ Frequency levels in a squared (left) and harmonic (right) effective potential for photons. $b)$ PhC cavity design creating the harmonic potential, the first rows of holes with period $a'$ is in blue. $c)$ False-color maps of the fundamental mode ($|H_z|^2$) calculated by FDTD along the main symmetry axis and and its representation in the reciprocal space (inset); the color scale is logarithmic and the same for all maps. $d)$ Extracted field amplitudes along the main axis ($y=0$, $z=0$) with vertical offset corresponding to the resonance $\nu_m$ and solutions (solid gray line) of the harmonic potential (dashed red line). $e)$ Corresponding frequency spacing and linear fit ($\Delta^2\nu m + \Delta\nu$); the estimated numerical error is less than $10 GHz$.} 
	\label{fig:CavityHarmonicPotential}
\end{figure}
%
The design of our resonator with a harmonic photonic potential revolves around the concept of the \textit{bi-chromatic} photonic lattice recently introduced by Alpeggiani et al.\cite{Alpeggiani_APL2015} in the context of Photonic Crystals. It results from the superposition of two periodic lattices with periods $a$ and $a^\prime$, e.g.: $V(x)=V_0\cos(2\pi x/a)+V_1\cos(2\pi x/a')$. This system provides a non trivial localization mechanism~\cite{albert_localization_2010} which has been investigated in the context matter waves~\cite{roati_anderson_2008}. Vanishingly small radiative losses have been predicted in photonic crystal cavities based on a bichromatic lattice, while the mode volume is still close to the diffraction limit~\cite{Alpeggiani_APL2015}. A crucial insight is the nearly perfect Gaussian envelope of the fundamental mode, which is consistent with an effective harmonic photonic potential.\\
In contrast to Refs.~\cite{Alpeggiani_APL2015,Simbula2016}, where the cavity is created by changing the period of a line of holes (and by reducing their radius), here we modify a missing line defect waveguide by setting the period of the first rows of holes to $a'=0.98a$, with $a=485$ nm (Fig.\ref{fig:CavityHarmonicPotential}b). The two rows are further displaced inwards by $\Delta s=0.015\sqrt{3}a$. The PhC structure is based on a lattice of holes with radius $r=0.27a$ etched in a 180 nm thick slab of $Ga_{0.5}In_{0.5}P$\cite{combrie2009high,martin2016photonic} with refractive index 3.17. The modes of the resonator are calculated by the Finite Difference in Time Domain method with resolution up to $a/50$ and sub-pixel-smoothing ensuring an absolute accuracy of about 100 GHz and a relative accuracy (frequency difference) of a few GHz.\\
The field distribution of the fundamental mode (Fig.\ref{fig:CavityHarmonicPotential}c) follows a Gaussian envelope along the main axis $x$. This is consistent with the representation of the mode in the reciprocal space; there the normalized amplitude drops below the -60 dB level in the light circle, implying that radiation losses are extremely small~\cite{akahane2003high,srinivasan_momentum_2002}. Consistently, the calculated Q factor is well above $10^7$. There are at least 6 other higher order modes forming a regular comb, spaced by $\Delta\nu\approx$ 450 GHz (Fig~\ref{fig:CavityHarmonicPotential}e). The envelope (Fig~\ref{fig:CavityHarmonicPotential}d) of these modes ($|H_z(x,0,0)|$ along the axis of symmetry is compared with the well known\cite{cohen_tannoudji} solution (Hermite-Gauss functions) of a free electron in a harmonic potential $V(x)=\nu_0+\Delta\nu/2 + \alpha\Delta \nu x^2/2$, with $\alpha=0.25\mu m^{-2}$. The agreement of both the calculated resonances and fields is almost perfect, therefore the PhC structure implements a harmonic photonic potential.\\  
We have calculated the eigenvalues of the Aubry-Andr\'{e}\cite{aubry_serge_analyticity_1980} or Harper Hamiltonian for the bichromatic potential~\cite{albert_localization_2010,modugno_exponential_2009}. It is found that the eigenvalues are not evenly spaced, when the potential takes the simple form $\omega^2+\Delta\cos(2\pi\beta l)$, as proposed in Ref.\cite{Alpeggiani_APL2015} ; rather, their spacing is constantly decreasing with increasing mode order ($\Delta^2\nu<0$). This approximation does not explain the effective harmonic potential. We speculate that its origin is intimately related to the structure of the photonic crystal. Also, we note that the tight binding (TB) approximation underlying this model might not be adequate. Interestingly, it was pointed out that the TB model is not adequate to describe coupled photonic crystal cavities\cite{lian_dispersion_2015}. This point is further discussed in the \textit{Appendix}.\\ 
Changing the commensurability parameter $\beta=a'/a$ primarily affects the volume of the modes, as discussed in ref.~\cite{Alpeggiani_APL2015}, but can also be used to control the mode spacing, which is a desirable feature. In other words, the harmonic potential seems to be a rather robust feature of the bichromatic design, as will be apparent in the next section discussing experiments. 
\section{Experiment}
\begin{figure}[htbp]
	\centering
	\includegraphics[width=\linewidth]{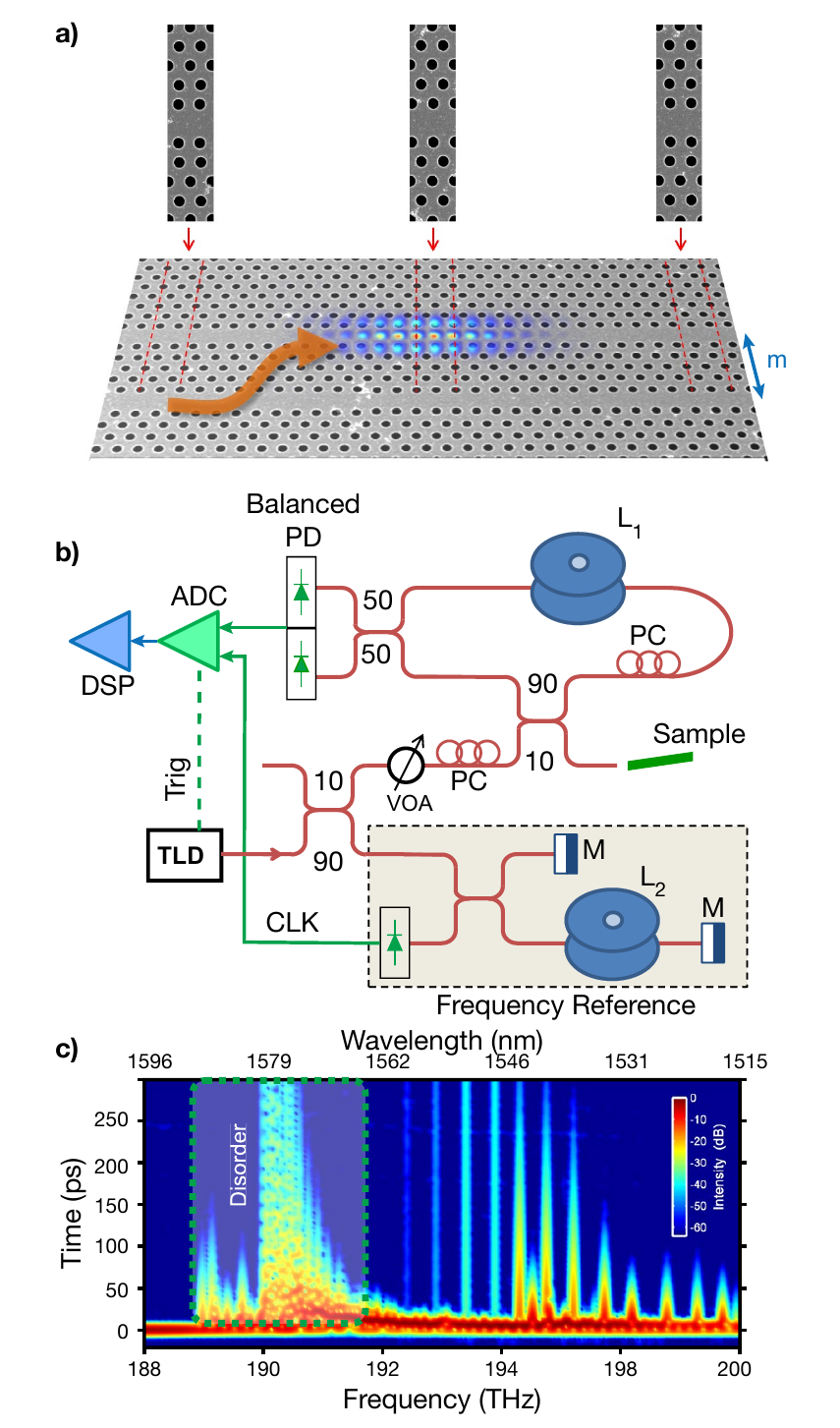}
	\caption{$a)$ SEM picture of the PhC resonator with input waveguide separated by $m=6$ rows and close-up views revealing the bichromatic design. The field distribution ($|E^2|$) of the fundamental mode is superimposed. $b)$ Measurement of the complex reflectance using OCT (Optical Coherent Tomography). PC: Polarization Controller, M: Mirror, TLD: Tunable Laser Diode, DUT: Device Under Test, VOA: Variable Optical Attenuator. The shaded area represents the unbalanced reference interferometer with $L_2$ the long arm $c)$ Time-Frequency (spectrogram) map of the reflected signal. The scale is logarithmic and set between -60 and 0 dB.} 
	\label{fig:Experiment}
\end{figure}

Two nominally identical copies ($A$ and $B$) of a chip have been fabricated using a standard process for III-V semiconductor alloys. The radius of the holes is varied according to the law $r = (0.0025l+0.27)a$ with $l=-10,-9,..,6$. The resonators are side-coupled to a single-ended waveguide spaced by $m=6$ or $m=5$ rows from the cavity, as shown in Fig.~\ref{fig:Experiment}a. The coupling waveguide is throttled on the right hand side in order to prevent further propagation. A total of 68 resonators, all functional, have been obtained.
%
%
%
The detection of high-Q resonances is not trivial. The straightforward procedure is to observe the light scattered out of the cavity with an infrared camera or a detector~\cite{sokolov_measurement_2016} and detecting the peaks. Care must be paid to avoid unwanted scattered light, while sensitivity might be an issue if the power level has to be limited to avoid nonlinear effects. Sensitivity is improved using optical homodyne detection~\cite{Simbula2016}.\\
Here, we analyze the complex reflectivity amplitude from the input waveguide of the device. This approach has some desirable peculiarities: first, it only requires access from the input channel, which is designed for a low-loss ($<3dB$, owing to an inverse taper\cite{tran2009photonic}) coupling to a fiber. Moreover, the simultaneous fitting of the phase and the amplitude with a zero/pole function allows the extraction of both the loaded and the intrinsic Q-factors while the phase signal provides an unambiguous discrimination of the under, over and critical coupling regimes. This will be discussed in the next section.\\ 
The complex reflectivity amplitude can be obtained using Optical Low Coherence Reflectometry (OLCR), which has been used to characterize photonic devices and ring resonators in particular~\cite{gottesman2004new,sanogo2013phase}. We used OLCR to measure the dispersion and disorder in PhC structures~\cite{parini2008time,combrie2007investigation}. Here we use Optical Coherent Tomography (OCT), which is well known in medical imaging~\cite{huang_optical_1991}). OCT is based on a continuously swept laser source, instead of a partially coherent source, which removes the need of a movable mirror. This also implies a much larger spectral power density which is particularly useful for narrow linewidth measurement.
The absolute frequency accuracy of the laser used here (Santec TSL 510C) is better than 1 GHz. As shown in Fig.~\ref{fig:Experiment}b, the instantaneous wavelength during the sweep is measured using a reference Michelson interferometer (shaded area) which is here made unbalanced using $L_2=5$ m of single mode fiber (SMF). \\
Consistently with a reflection measurement, the main interferometer is arranged in the Michelson configuration with the sample coupled at the end of one arm. The interference signal is detected by a balanced photodetector (Thorlabs PDB450c) and then acquired with a ADC card (AlazarTech 460) triggered by the signal generated by the reference interferometer. This ensures that the data samples are evenly spaced in frequency. We operated our own OCT system with a sweep time of a few seconds over the C+L bands (1510 nm to 1630 nm) and obtained a spectral resolution of about 20 MHz.\\
A time-frequency spectrogram (Fig.\ref{fig:Experiment}c) is obtained by applying a sliding filter with a suitable window to the interferogram and then by Fourier transforming~\cite{parini2008time,gottesman2010time}. This reveals the arrival time of the light after propagation in the input waveguide from which the dispersion can be identified and extracted\cite{sancho2012integrable,colman2012blue}. It also reveals the disorder-induced scattering. Finally, the cavity resonances are clearly visible as long decay lines. In Fig.\ref{fig:Experiment}c, more than 10 resonances are apparent and clearly distinguishable from accidental resonances due to disorder near the waveguide band edge (189-192 THz). The power level used in this experiment ($\approx 100$ nW) is low enough to avoid any nonlinear effect in the cavities, including thermo-optic effects. Nonetheless, the dynamic range remains remarkably large ($>> 60$ dB). 
\section{Spectroscopy using complex reflection spectrum}
%
%
%
\begin{figure}[htbp]
	\centering
	\includegraphics[width=\linewidth]{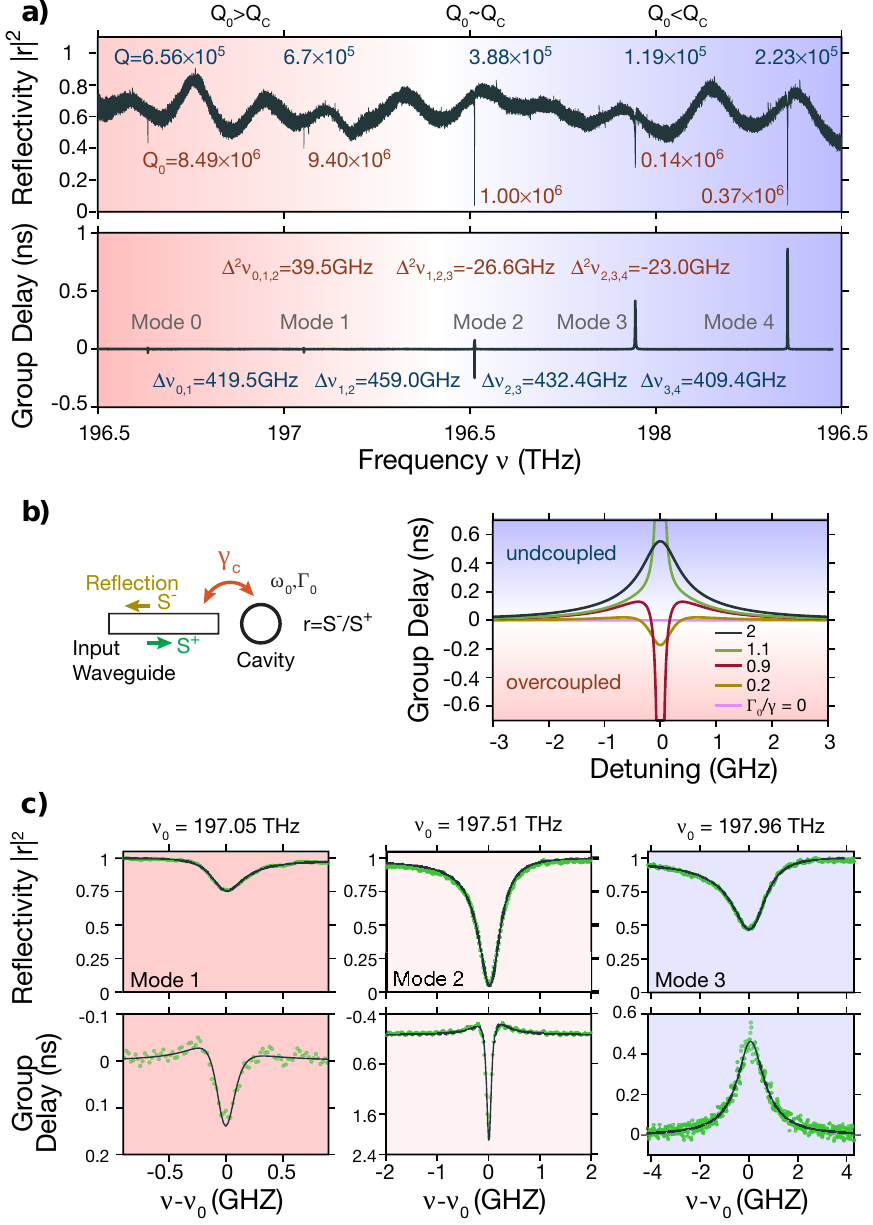}
	\caption{$a)$ Measured complex reflectivity (power and group delay) spectra extracted from the interferogram of the cavity with $\Delta r=0.015a$ and $m=6$. $b)$ cavity coupled to a single-ended waveguide and related zero-pole model for the group delay as a function of coupling $\gamma/\Gamma_0$. $c)$ Close-up of the measured complex spectrum near three resonances corresponding to strongly over-coupled, critically coupled and under-coupled resonator, and corresponding fit with the pole-zero model.}
	\label{fig:Complex_Spectrum}
\end{figure}
%
The complex reflection spectrum $r(\nu)$ is shown in Fig~\ref{fig:Complex_Spectrum}a for a representative resonator ($\Delta r=0.015a$ and $m=6$). Resonances appear as dips in the amplitude signal and as peaks or dips in the group delay. In contrast to the amplitude signal, the group delay is virtually zero outside the resonances, therefore making their identification unambiguous. Moreover, the lineshapes of the group delay are symmetric, ruling nonlinear effects out. The spectral separation (Free Spectral Range) $\Delta\nu$ between the modes is about 430 GHz. The measured second order difference $\Delta^2\nu_j = \nu_{i+1}+\nu_{i-1}-2\nu_{i}$ corresponds to the frequency mismatch for degenerate four wave mixing.\\
The Q-factors associated to internal losses ($Q_0=\omega/\Gamma_0$) and coupling ($Q_c=\omega/\gamma$)   are extracted (Fig. \ref{fig:Complex_Spectrum}b) by fitting the amplitude and the group delay near the peaks using a pole-zero function $r(\nu)  = \frac{2 \pi \nu-z}{2 \pi\nu-p}$.\\
Following the fundamental mode, four other resonances are found, all with loaded Q factor ranging from $10^5$ to about $7\times 10^5$. The estimated intrinsic Q factor for the three first resonances is in the $10^6-10^7$ range. Let us consider three representative cases: over-coupled resonances (modes 0 and 1), close to critical coupling (mode 2) and undercoupled (mode 3). In all cases, the amplitude and the group delay are very well approximated by the zero-pole function. Moreover, it is apparent that in the case of strongly over-coupled modes, the group delay is negative and small (modes 0 and 1), it increases near the critical coupling (mode 2) and changes sign when becoming under-coupled (modes 3 and 4), exactly as expected from theory (Fig~\ref{fig:Complex_Spectrum}b.\\  
%
%
\begin{figure}[htbp]
	\centering
	\includegraphics[width=\linewidth]{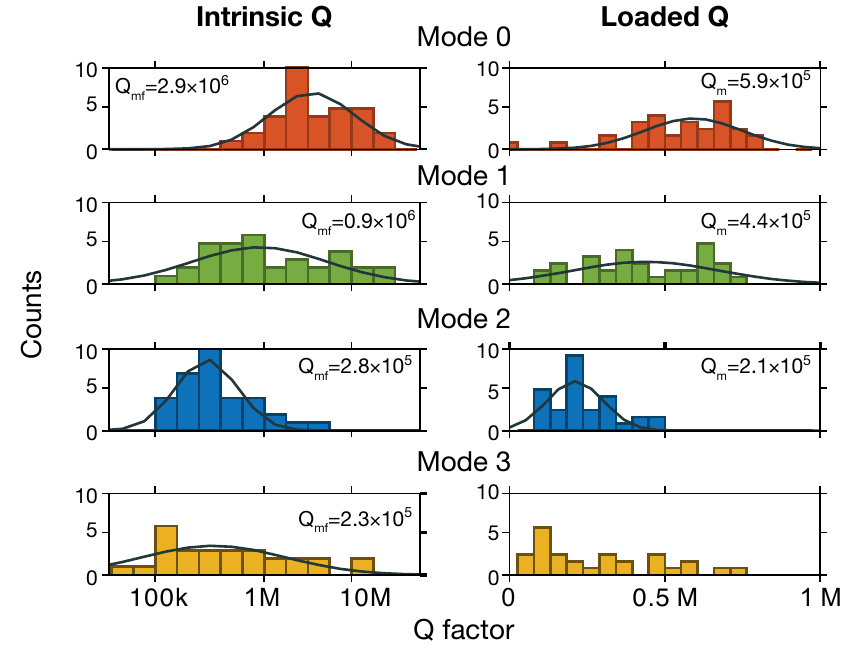}
	\caption{Left (right) column: measured intrinsic (loaded) Q-factors on the 34 resonators with $m=6$.  Histograms for the occurrences of Q for mode 0 to mode 3 and log-normal (normal) fit for the distributions of the intrinsic (loaded) Q factors.}
	\label{fig:Qfactors}
\end{figure}
%
A very important question is whether such large Q factors are accidental or reproducible. Therefore, measurements have been performed over 34 resonators and summarized by histograms in Fig.~\ref{fig:Qfactors}
describing the occurrences of the Q factors for the first four modes. A \textit{log-normal} distribution $P=\exp(-[\log(Q/Q_mf)]^2/\sigma^2_Q)$ is apparent for the measured intrinsic Q factors, with the most frequent value $Q_{mf}$ decreasing from $2.9\times10^6$ for the fundamental order mode to $3\times10^5$ for the fourth order mode. The calculated Q factors for the uncoupled waveguides are well above $10^7$ for the first 4 modes and for any of the values of the radius of the holes considered here. Thus, intrinsic radiative losses play no role here. On the other hand, we estimate residual~\cite{martin2017gainp} and nonlinear absorption to be negligible up to Q factors of several millions. Thus, the statistics measured here should be dominated by disorder.\\
The understanding of disorder in two dimensional photonic crystals has progressed remarkably in the last years owing to several theoretical\cite{savona_electromagnetic_2011,mann_theory_2015,thyrrestrup_statistical_2012} and experimental studies, particularly focused on the mechanisms of Anderson localisation~\cite{john_strong_1987} in disordered waveguides~\cite{topolancik_experimental_2007,smolka_probing_2011,faggiani_lower_2016,xue_threshold_2016}. Based on the calculations~\cite{minkov_statistics_2013} for heterostructure cavities and related experiments\cite{taguchi_statistical_2011}, we estimate the roughness $\sigma_d \approx 0.002a\approx 1$ nm, ($\sigma_d$ being the magnitude of the random fluctuations), which is slightly larger than in Silicon. Very recently, comparable Q factors have been reported on a bichromatic PhC cavity made of Silicon~\cite{Simbula2016} based on a different design, where the overlap of the field with the holes is stronger. This might have an impact on the measured Q factor.\\
\indent Existing literature provides some insights to understand the origin of the log-normal distribution of the intrinsic Q factor. First, such distribution is predicted in Ref.~\cite{thyrrestrup_statistical_2012} for spontaneously localized modes in disordered waveguides. Moreover, it has been shown that these modes extend over a minimum of $6\mu$m \cite{faggiani_lower_2016}, which is reasonably close to the size of the modes of the harmonic potential (Fig.~\ref{fig:CavityHarmonicPotential}c); on the other hand, a dependence of the disorder induced radiative leakage on the spatial size of the mode has been observed\cite{xue_threshold_2016}. Therefore, we expect that our intentionally localized modes inherit similar statistical features regarding disorder induced losses.
We conjecture that considering the intrinsic Q factor removes the dominant contribution of the deterministic loss channel which is the input waveguide, and therefore this is close to the situation considered in Ref.~\cite{thyrrestrup_statistical_2012}. It is worth noting that a log-normal distribution also describes the statistics of the emission rates of colloidal quantum dots randomly placed inside a $TiO_2$ inverse opal\cite{van_driel_statistical_2007}.\\
The loaded Q factor, in contrast, follows a normal distribution, and it appears insensitive of the radius of the holes, as expected from FDTD calculations. This is in line with Refs.~\cite{taguchi_statistical_2011,minkov_statistics_2013}. Given the peculiar spatial structure of higher order modes in a harmonic potential, it is less straightforward to apply the theory above to explain the observed dependence of the intrinsic Q on the mode order, while theory predicts very large Q for all these modes in absence of disorder. Also the origin of the very different spread of the intrinsic Q factors with the mode order is unclear. A possible hint might come from  Ref.~\cite{xue_threshold_2016}.
\section{Discussion: Frequency Comb in PhC}
%
%
\begin{figure}[htbp]
	\centering
	\includegraphics[width=\linewidth]{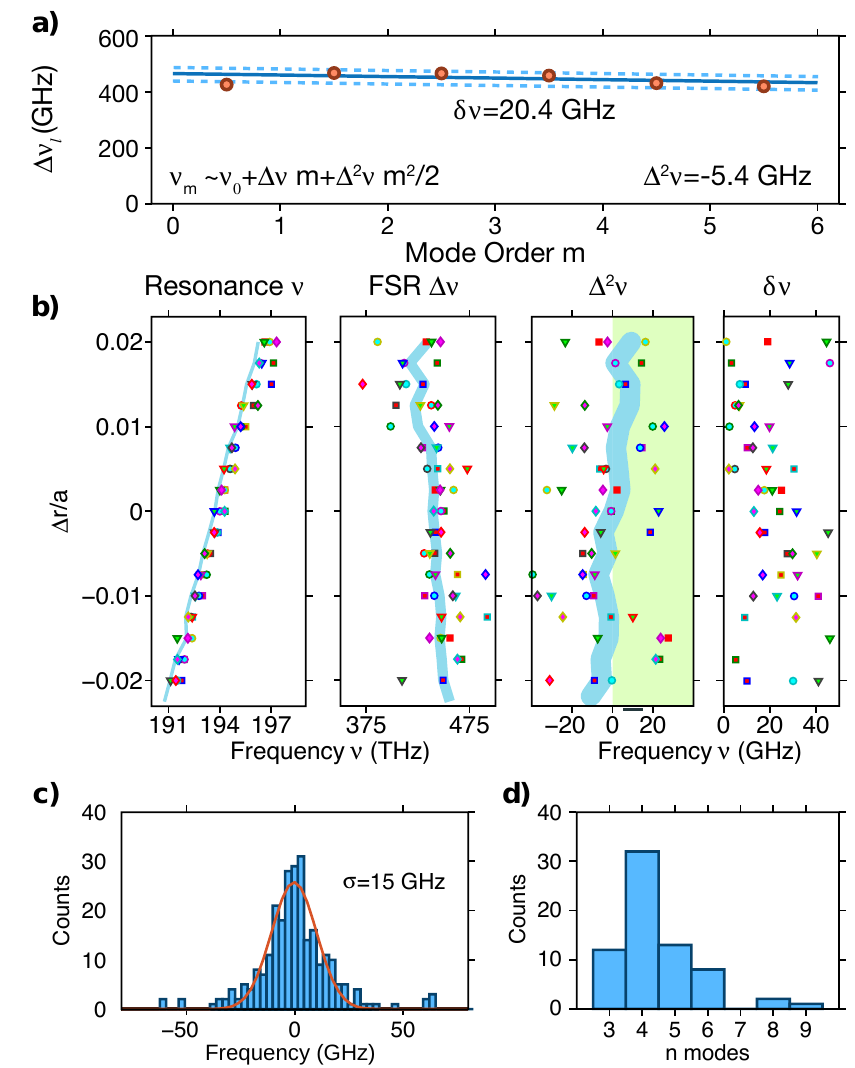}
	\caption{$a)$ Measured FSR ($\Delta\nu_i$) vs. mode order for cavity with $\Delta r=0$, sample $B$, distance $m=6$, and polynomial fit (solid line) with norm of the residual $\delta\nu=|\nu_m-p(m)|$ (dashed line). $b)$ Measured coefficients of the interpolating polynomial and residuals as a function of the radius of the holes $\Delta r$. Colored symbols denote two nominally identically samples, $A$ and $B$, and two waveguide to cavity distances ($m=6$) and ($m=5$). The shaded area represents anomalous dispersion. FDTD calculations (cyan line) with blur representative of the numerical error. Histograms of the number of modes $c)$ with $Q>10^4$  and of the residuals $d)$, fitted with a normal distribution (red line) with standard deviation $\sigma$.}
	\label{fig:statistics_comb}
\end{figure}
%
We now analyse the measured spacing of the resonances with respect to the ideal case of a frequency comb and to what extent it is possible to control it, in analogy to what is made with other kind of resonators.
Let us first consider, as an example, a cavity with parameters $\Delta r=0$ and $m=6$ (Fig.~\ref{fig:statistics_comb}a). The seven modes, with intrinsic Q factor up to $3\times10^6$, are fitted with a second order polynomial $\nu_m=p(m)=\nu_0+\Delta\nu m + \Delta^2\nu/2 m^2$. This defines the frequency $\nu_0$ of the fundamental order mode, the free spectral spacing $\Delta\nu$, the dispersion $\Delta^2\nu$ and the norm of the residuals $\delta\nu=|\nu_m-p(m)|$ of the fit. This measurement has been performed over all the 68 fabricated resonators and the result is summarized in Fig.~\ref{fig:statistics_comb}b. $\delta\nu$ varies over the samples and in some cases is close to a few GHz. Importantly, the distribution of $\delta\nu$ clearly follows a normal distribution (Fig.~\ref{fig:statistics_comb}c). Therefore, we ascribe the deviation of the resonances from the fitting potential to the structural disorder. This also indicates that the parameters extracted through a fitting polynomial are much less sensitive to the fabrication disorder and therefore reflect the underlying dispersion of the resonator.\\
It is interesting to compare with the statistical measurements performed on individual resonators made with arguably the best fabrication process for PhCs~\cite{taguchi_statistical_2011}, where a standard deviation of about 40 GHz is reported. The smaller width $\sigma$ = 15 GHz of the distribution in Fig.~\ref{fig:statistics_comb}c is explained by the fact that $\delta\nu$ describes frequency intervals within the same resonator, hence correlated. On the other hand, the peculiar distribution of the modes of the harmonic potential, overlapping partially, suggests that these correlation are not as strong as in other resonators.\\
A clear trend is observed for $\nu_0$ and $\Delta\nu$ in Fig.~\ref{fig:statistics_comb}b, which is in very good agreement with FDTD calculations. Subtle features are guessed from the correlated deviations of the experimental points. This is even more visible in the plot of the dispersion. We conjecture that these features, not predicted by calculations, could be related to Moir\'{e} effects due to the finite resolution of the e-beam.\\
The most important point is that the fitted parameter $|\Delta^2\nu|$ is in general smaller than 20 GHz, with several data points much closer to zero and small residuals $\delta\nu$ as well. Considering, instead, the direct measurements of $\Delta^2\nu_j$, there are a few experimental points as small as 1 GHz. This corresponds very well with the calculated dispersion (blurred line in Fig.~\ref{fig:statistics_comb}b) crossing the zero as the radius of the holes is changed. Therefore, at some point, $\Delta^2\nu$ turns slightly positive which would correspond to anomalous dispersion in a ring resonator, hence achieving the favorable configuration for parametric oscillation and comb generation. Finally, most of the resonators have four high-Q modes, but more modes are not rare (Fig.~\ref{fig:statistics_comb}d).\\
From a more practical point of view, a control on the resonances is highly desirable and only requires a spectral displacement of about ten GHz, which could be obtained by a local temperature change of about 1 K. This should be fairly easy, given the peculiar localization of higher modes (Fig~\ref{fig:CavityHarmonicPotential}). The individual control of the modes in a chain of coupled PhC resonators issued from the same technology ($GaInP$) has been demonstrated over a much broader tuning range using holographic techniques to induce a well controlled thermal gradient~\cite{sokolov_tuning_2016}. An implementation of the concept more appropriate for applications would exploit the Joule effect in close electric paths~\cite{han2015high}.\\
\section{Conclusion}
Through the concept of bichromatic lattice, we have implemented an effective harmonic potential in a photonic crystal resonator and observed the clear signature of ``comb'' of equi-spaced resonances, starting from the fundamental order mode. This is fundamentally different from Fabry-Perot, ring resonators or whispering gallery modes, which are highly over-moded. Measurements have been performed using the Optical Coherent Tomography, which gives access to the complex amplitude of the reflected signal. We show an accurate and reliable measurement of the frequency spacing and of both the loaded and the intrinsic Q factors. The statistical analysis, carried over 68 resonators, concluded into a log-normal distribution of the intrinsic Q factor, with most frequent value close to 3 million in a III-V semiconductor structure. The dispersion of the resonator is extracted through a polynomial fit, with residuals following a normal distribution. \\
The main result is that the dispersion is small and crosses zero as the radius of the holes is changed. Many structures with very high-Q resonances and close to ideal frequency spacing are found. Also considering that the mode volume is about $0.9(\lambda/n)^3$, the power threshold $P_{th}$ for parametric oscillation should be quite small. Based on a calculated nonlinear interaction volume $V_{FWM}\approx 25(\lambda/n)^3$ and using the formula: $P_{th}\approx\frac{\varepsilon_r V_{FWM}\omega}{4 c_0 n_2Q^2}$ valid for overcoupled pump  and the average loaded Q for the first 3 modes in Fig.~\ref{fig:Qfactors} leads to an prediction of $P_{th}\approx 30\mu W$. Moreover, it is expected that nonlinearity would compensate a residual positive mismatch $\Delta^2\nu$, a mechanism which is well known in high-Q resonators~\cite{kippenberg_kerr-nonlinearity_2004}. Our prediction for the threshold of parametric oscillation is to be compared with the state of the art, e.g. 300 $\mu W$ in $R\approx100$ $\mu$m $SiN$ ring resonators~\cite{ji_breaking_2016} or 3 mW in $AlGaAs$ devices\cite{pu_efficient_2016}. This low threshold is remarkable also considering that the nonlinearity $n_2$ of $GaInP$ is about half that of $AlGaAs$.
\section*{Funding Information}
This work was funded by the Minist\`ere de l'Enseignement Sup\'erieur et de la Recherche  through the contact ``CONDOR'' (Flagship NanoSaclay). A.D.R, S.C. and G.L. acknowledge support from the ERC grant ``PHAROS'' (P.I.  A.P. Mosk, g.a. 279248) and the FET Proactive project ``HOT'' (g.a. 732894)
\section*{Acknowledgments}
The authors thank O. Parillaud from III-VLab for the MOCVD growth of the epitaxy, D. Thenot for PECVD deposition and S. Xavier for the e-beam lithography. We thanks W.L. Vos and A. P. Mosk for enlightening discussions on disorder.\\

\section*{APPENDIX: The Harper Andr\'{e}-Aubry model and the effective Photonic Harmonic Potential}
\begin{figure}[htbp]
\includegraphics[width=\columnwidth]{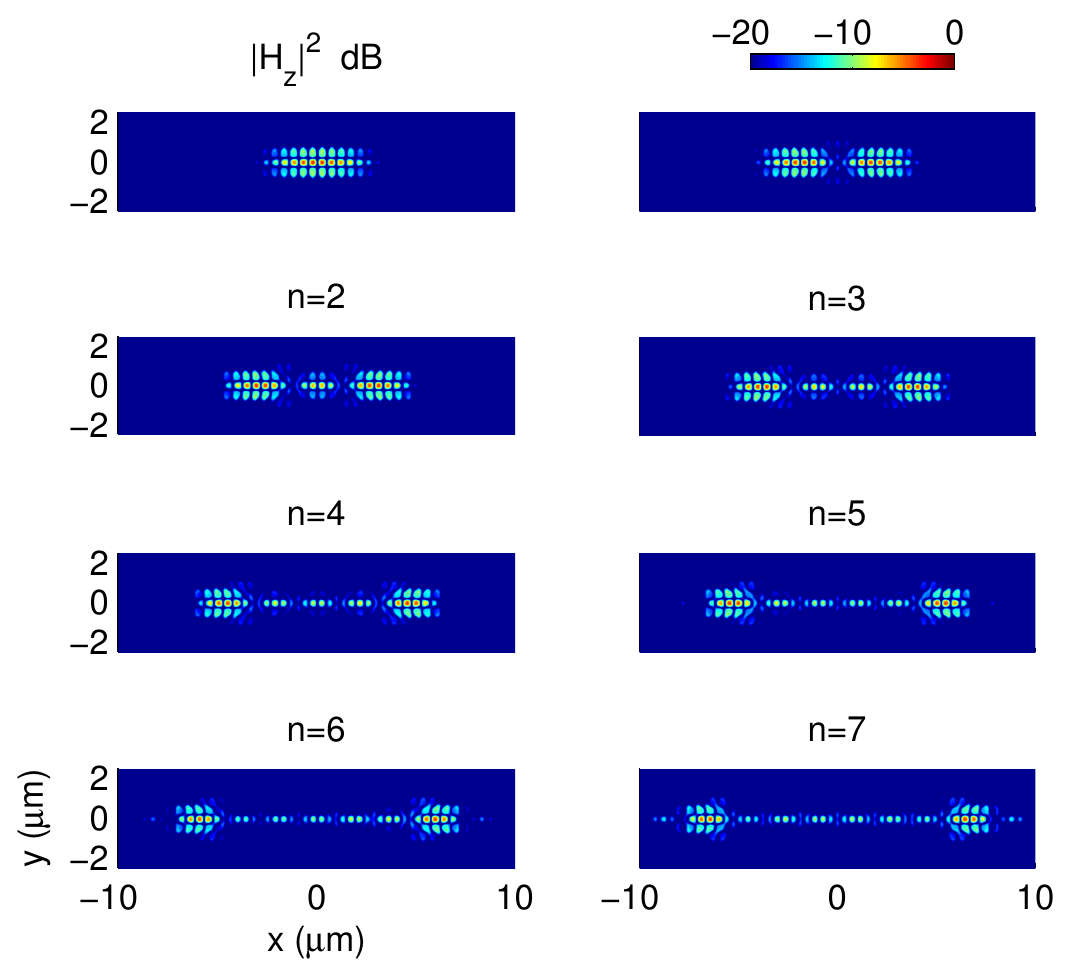}
\caption{Calculated modes from order 0 to order 7 for $\Delta r$=0 with resolution $\Delta x=a/28$. False colour map of $|H_z|^2$ at the plane of symmetry $z=0$ in logarithmic scale.}
\label{fig:modes_th}
\end{figure}
Figure \ref{fig:modes_th} represents the field distributions of the first 8 modes calculated by FDTD, which relate to Fig~\ref{fig:CavityHarmonicPotential}c. The envelopes reveal the characteristic distribution of a harmonic oscillator, where the field distribution takes the role of the density of the probability.\\ 
Alpeggiani et al.\cite{Alpeggiani_APL2015} have derived a Tight-Binding (TB) model of a Photonic Crystal cavity with a bichromatic lattice: 
\begin{equation}
\omega_l^2 c_l - J_l[c_{l-1} +c_{l+1}]=4\pi^2\nu^2 c_l
\end{equation}
here $c_l$ describes the amplitude of the Wannier wavefunctions on each site $l$ of the discretised structure. 
\begin{figure}[htbp]
\includegraphics[width=\columnwidth]{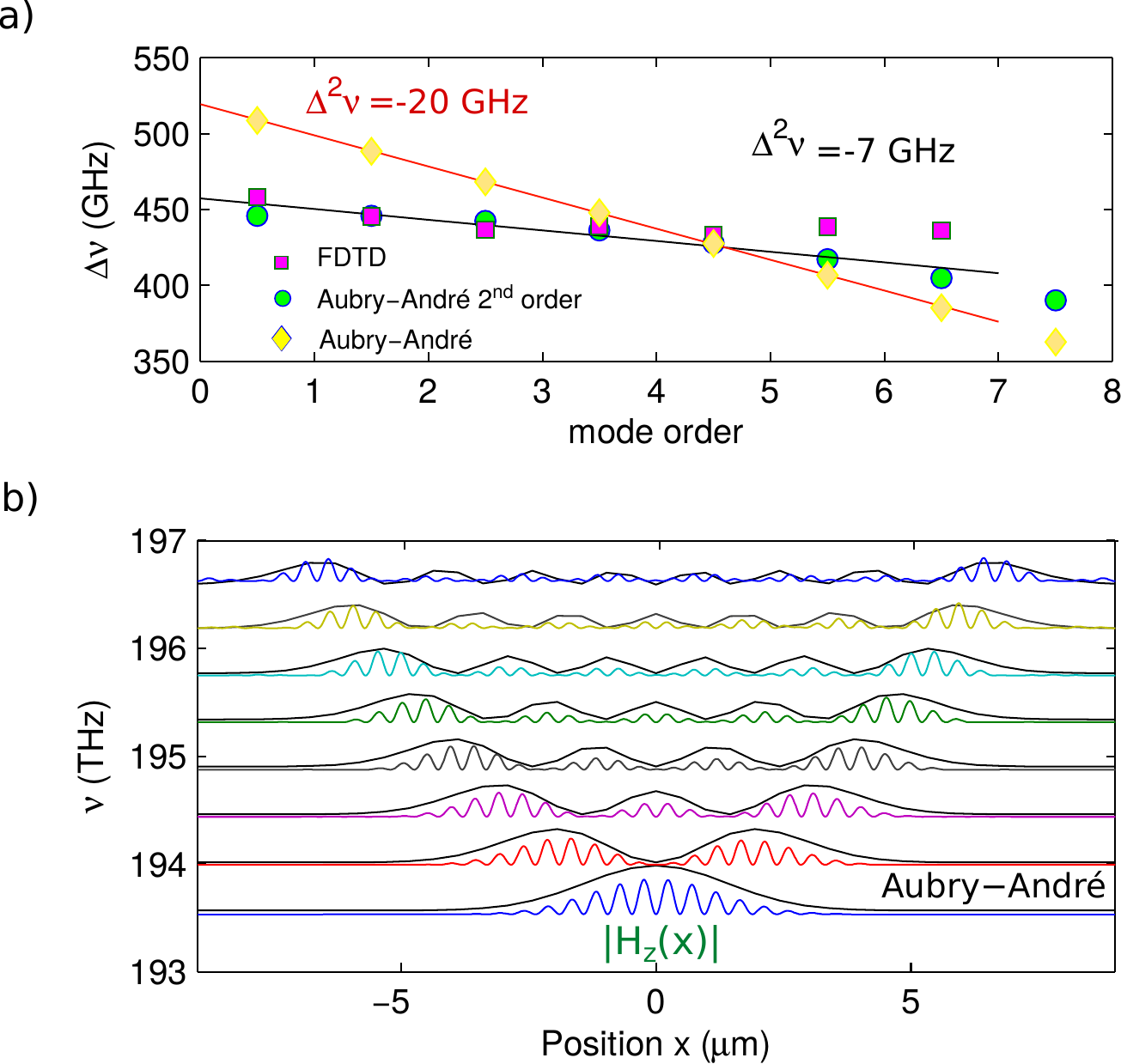}
\caption{Comparison between the modes of the PhC bichormatic cavity calculated by FDTD and the eigensolution of the Aubry-Andr\'{e} potential. $a)$ Frequency intervals of the eigenvalues $\nu_j$ and $b)$ corresponding field envelopes with offset $\nu_j$ calculated by FDTD at $H_z(x,0,0)$ (coloured solid) and from the Aubry-Andr\'{e} (black lines) model.}
\label{fig:aubry_potential}
\end{figure}
There, it is also shown that this model reduces to the Harper / Andr\'{e}-Aubry Hamiltonian which describes a bichromatic potential \cite{albert_localization_2010,modugno_exponential_2009}. It is assumed that the hopping parameter is constant $J_j=J$ and that the potential is approximated by: $\omega^2_l\approx \omega_0^2+\Delta\cos(2\pi \beta l)$, with $\beta=a'/a$ the commensurability parameter.\\
We now apply this model to the photonic crystal structure considered in our work. The first 8 eigensolutions $\nu$ are shown in Fig.~\ref{fig:aubry_potential}. The parameters of the model are chosen to fit the field distributions corresponding to \ref{fig:modes_th} and also shown in Fig. 1c of the main article. This yields $J=0.0036(2\pi c_0/a)^2$, $\Delta/J=0.7$ and $\nu_0=202.7$ THz.\\
As shown in Fig.~\ref{fig:aubry_potential}a, the eigenfrequencies are not equispaced, the spacing decreasing monotonically by about $\Delta^2\nu$=20 GHz. The approximation of a sinusoidal potential is clearly not adequate to account for the flat dispersion $\Delta^2\nu\rightarrow 0$ calculated for the PhC cavity. \\
A better match is obtained by adding a second order correction (with inverse period $2\beta$) to the potential: $\omega^2_l\approx \omega_0^2+\Delta\cos(2\pi l \beta)+\Delta_2\cos(4\pi l \beta)$, with $\nu_0=203THz$, $\Delta/J=0.7$, $J=0.038(2\pi c_0/a)^2$ and $\Delta_2/J=0.065$.
Therefore, the harmonic photonic effective potential demonstrated in bichromatic PhC does not seem to stem from a general property of the Harper / Andr\'{e}-Aubry Hamiltonian, rather, it is seemingly an unexpected property of PhC crystals with the bichromatic design. What is remarkable here is that this property results from a design which is entirely described by a few parameters: $\beta$, the radius of the holes of the PhC lattice, the refractive index and the thickness of the dielectric slab. Moreover, the harmonic potential seems only to depend on the index contrast, as the other parameters have a small influence on the dispersion. \\


%
\bibliography{2017_Combrie_MultiMode.bbl}

\begin{thebibliography}{10}
\newcommand{\enquote}[1]{``#1''}

\bibitem{kippenberg_kerr-nonlinearity_2004}
T.~J. Kippenberg, S.~M. Spillane, and K.~J. Vahala,
  \enquote{Kerr-{Nonlinearity} {Optical} {Parametric} {Oscillation} in an
  {Ultrahigh}- {Q} {Toroid} {Microcavity},} Physical Review Letters \textbf{93}
  (2004).

\bibitem{matsko_review_2005}
A.~B. Matsko, A.~A. Savchenkov, D.~Strekalov, V.~S. Ilchenko, and L.~Maleki,
  \enquote{Review of applications of whispering-gallery mode resonators in
  photonics and nonlinear optics,} IPN Progress Report \textbf{42}, 1--51
  (2005).

\bibitem{grassani_micrometer-scale_2015}
D.~Grassani, S.~Azzini, M.~Liscidini, M.~Galli, M.~J. Strain, M.~Sorel, J.~E.
  Sipe, and D.~Bajoni, \enquote{Micrometer-scale integrated silicon source of
  time-energy entangled photons,} Optica \textbf{2}, 88 (2015).

\bibitem{preble_-chip_2015}
S.~F. Preble, M.~L. Fanto, J.~A. Steidle, C.~C. Tison, G.~A. Howland, Z.~Wang,
  and P.~M. Alsing, \enquote{On-{Chip} {Quantum} {Interference} from a {Single}
  {Silicon} {Ring}-{Resonator} {Source},} Physical Review Applied \textbf{4}
  (2015).

\bibitem{reimer_generation_2016}
C.~Reimer, M.~Kues, P.~Roztocki, B.~Wetzel, F.~Grazioso, B.~E. Little, S.~T.
  Chu, T.~Johnston, Y.~Bromberg, L.~Caspani, D.~J. Moss, and R.~Morandotti,
  \enquote{Generation of multiphoton entangled quantum states by means of
  integrated frequency combs,} Science \textbf{351}, 1176--1180 (2016).

\bibitem{li_efficient_2016}
Q.~Li, M.~Davanço, and K.~Srinivasan, \enquote{Efficient and low-noise
  single-photon-level frequency conversion interfaces using silicon
  nanophotonics,} Nature Photonics \textbf{10}, 406--414 (2016).

\bibitem{kippenberg_microresonator-based_2011}
T.~J. Kippenberg, R.~Holzwarth, and S.~A. Diddams,
  \enquote{Microresonator-{Based} {Optical} {Frequency} {Combs},} Science
  \textbf{332}, 555--559 (2011).

\bibitem{huang_broadband_2016}
S.-W. Huang, J.~Yang, M.~Yu, B.~H. McGuyer, D.-L. Kwong, T.~Zelevinsky, and
  C.~W. Wong, \enquote{A broadband chip-scale optical frequency synthesizer at
  2.7 x 10-16 relative uncertainty,} Science Advances \textbf{2},
  e1501489--e1501489 (2016).

\bibitem{ji_breaking_2016}
X.~Ji, F.~A. Barbosa, S.~P. Roberts, A.~Dutt, J.~Cardenas, Y.~Okawachi,
  A.~Bryant, A.~L. Gaeta, and M.~Lipson, \enquote{Breaking the {Loss}
  {Limitation} of {On}-chip {High}-confinement {Resonators},} arXiv preprint
  arXiv:1609.08699  (2016).

\bibitem{pu_efficient_2016}
M.~Pu, L.~Ottaviano, E.~Semenova, and K.~Yvind, \enquote{Efficient frequency
  comb generation in {AlGaAs}-on-insulator,} Optica \textbf{3}, 823 (2016).

\bibitem{xuan_high-q_2016}
Y.~Xuan, Y.~Liu, L.~T. Varghese, A.~J. Metcalf, X.~Xue, P.-H. Wang, K.~Han,
  J.~A. Jaramillo-Villegas, A.~Al~Noman, C.~Wang, S.~Kim, M.~Teng, Y.~J. Lee,
  B.~Niu, L.~Fan, J.~Wang, D.~E. Leaird, A.~M. Weiner, and M.~Qi,
  \enquote{High-{Q} silicon nitride microresonators exhibiting low-power
  frequency comb initiation,} Optica \textbf{3}, 1171 (2016).

\bibitem{john_strong_1987}
S.~John, \enquote{Strong localization of photons in certain disordered
  dielectric superlattices,} Physical Review Letters \textbf{58}, 2486--2489
  (1987).

\bibitem{notomi_theory_2000}
M.~Notomi, \enquote{Theory of light propagation in strongly modulated photonic
  crystals: {Refractionlike} behavior in the vicinity of the photonic band
  gap,} Physical Review B \textbf{62}, 10696 (2000).

\bibitem{painter_wannier-like_2003}
O.~Painter, K.~Srinivasan, and P.~E. Barclay, \enquote{Wannier-like equation
  for the resonant cavity modes of locally perturbed photonic crystals,}
  Physical Review B \textbf{68} (2003).

\bibitem{savona_electromagnetic_2011}
V.~Savona, \enquote{Electromagnetic modes of a disordered photonic crystal,}
  Physical Review B \textbf{83} (2011).

\bibitem{lalanne_photon_2008}
P.~Lalanne, C.~Sauvan, and J.~Hugonin, \enquote{Photon confinement in photonic
  crystal nanocavities,} Laser \& Photonics Review \textbf{2}, 514--526 (2008).

\bibitem{kim_characteristics_2004}
S.-H. Kim, G.-H. Kim, S.-K. Kim, H.-G. Park, Y.-H. Lee, and S.-B. Kim,
  \enquote{Characteristics of a stick waveguide resonator in a two-dimensional
  photonic crystal slab,} Journal of Applied Physics \textbf{95}, 411--416
  (2004).

\bibitem{barclay_nonlinear_2005}
P.~E. Barclay, K.~Srinivasan, and O.~Painter, \enquote{Nonlinear response of
  silicon photonic crystal micresonators excited via an integrated waveguide
  and fiber taper,} Optics Express \textbf{13}, 801 (2005).

\bibitem{srinivasan_momentum_2002}
K.~Srinivasan and O.~Painter, \enquote{Momentum space design of high-{Q}
  photonic crystal optical cavities,} Opt. Express, OE \textbf{10}, 670--684
  (2002).

\bibitem{tanaka_design_2008}
Y.~Tanaka, T.~Asano, and S.~Noda, \enquote{Design of {Photonic} {Crystal}
  {Nanocavity} {With} \${Q}\$-{Factor} of
  \$\{\{{\textbackslash}sim\}10{\textasciicircum}\{9\}\}\$,} Journal of
  Lightwave Technology \textbf{26}, 1532--1539 (2008).

\bibitem{notomi_ultrahigh-q_2008}
M.~Notomi, E.~Kuramochi, and H.~Taniyama, \enquote{Ultrahigh-{Q} nanocavity
  with 1d photonic gap,} Optics Express \textbf{16}, 11095--11102 (2008).

\bibitem{asano_photonic_2017}
T.~Asano, Y.~Ochi, Y.~Takahashi, K.~Kishimoto, and S.~Noda, \enquote{Photonic
  crystal nanocavity with a {Q} factor exceeding eleven million,} Optics
  Express \textbf{25}, 1769 (2017).

\bibitem{mccutcheon_design_2008}
M.~W. McCutcheon and M.~Loncar, \enquote{Design of a silicon nitride photonic
  crystal nanocavity with a {Quality} factor of one million for coupling to a
  diamond nanocrystal,} Optics Express \textbf{16}, 19136 (2008).

\bibitem{gong_photonic_2010}
Y.~Gong and J.~Vuckovi\'c, \enquote{Photonic crystal cavities in silicon
  dioxide,} Applied Physics Letters \textbf{96}, 031107 (2010).

\bibitem{bazin_design_2014}
A.~Bazin, R.~Raj, and F.~Raineri, \enquote{Design of {Silica} {Encapsulated}
  {High}-{Q} {Photonic} {Crystal} {Nanobeam} {Cavity},} Journal of Lightwave
  Technology \textbf{32}, 952--958 (2014).

\bibitem{eichenfield_optomechanical_2009}
M.~Eichenfield, J.~Chan, R.~M. Camacho, K.~J. Vahala, and O.~Painter,
  \enquote{Optomechanical crystals,} Nature \textbf{462}, 78--82 (2009).

\bibitem{quan_deterministic_2011}
Q.~Quan and M.~Loncar, \enquote{Deterministic design of wavelength scale,
  ultra-high {Q} photonic crystal nanobeam cavities,} Optics express
  \textbf{19}, 18529--18542 (2011).

\bibitem{Alpeggiani_APL2015}
F.~Alpeggiani, L.~C. Andreani, and D.~Gerace, \enquote{Effective bichromatic
  potential for ultra-high q-factor photonic crystal slab cavities,} Applied
  Physics Letters \textbf{107}, 261110 (2015).

\bibitem{albert_localization_2010}
M.~Albert and P.~Leboeuf, \enquote{Localization by bichromatic potentials
  versus {Anderson} localization,} Physical Review A \textbf{81} (2010).

\bibitem{roati_anderson_2008}
G.~Roati, C.~D{'} Errico, L.~Fallani, M.~Fattori, C.~Fort, M.~Zaccanti,
  G.~Modugno, M.~Modugno, and M.~Inguscio, \enquote{Anderson localization of a
  non-interacting {Bose-Einstein} condensate,} Nature \textbf{453},
  895--898 (2008).

\bibitem{Simbula2016}
A.~Simbula, M.~Schatzl, L.~Zagaglia, F.~Alpeggiani, L.~Andreani,
  F.~Sch{\"a}ffler, T.~Fromherz, M.~Galli, and D.~Gerace, \enquote{Realization
  of high-{Q/V} bichromatic photonic crystal cavities defined by an effective
  aubry-andr$\backslash$'e-harper potential,} APL Photonics \textbf{2}, 056102  (2017).

\bibitem{combrie2009high}
S.~Combri{\'e}, Q.~V. Tran, A.~De~Rossi, C.~Husko, and P.~Colman, \enquote{High
  quality {GaInP} nonlinear photonic crystals with minimized nonlinear
  absorption,} Applied Physics Letters \textbf{95}, 221108 (2009).

\bibitem{martin2016photonic}
A.~Martin, S.~Combri{\'e}, and A.~De~Rossi, \enquote{Photonic crystals
  waveguides based on wide-gap semiconductor alloys,} Journal of Optics
  (2016).

\bibitem{akahane2003high}
Y.~Akahane, T.~Asano, B.-S. Song, and S.~Noda, \enquote{High-q photonic
  nanocavity in a two-dimensional photonic crystal,} Nature \textbf{425},
  944--947 (2003).

\bibitem{cohen_tannoudji}
F.~L. C.~Cohen-Tannoudji, B.~Diu, \emph{M{\'e}chanique Quantique, Tome I}
  (Hermann, 1973).

\bibitem{aubry_serge_analyticity_1980}
S.~Aubry and A.~Gilles, \enquote{Analyticity breaking and {Anderson}
  localization in incommensurate lattices,} Ann. Israel Phys. Soc. \textbf{3}.

\bibitem{modugno_exponential_2009}
M.~Modugno, \enquote{Exponential localization in one-dimensional quasi-periodic
  optical lattices,} New Journal of Physics \textbf{11}, 033023 (2009).

\bibitem{lian_dispersion_2015}
J.~Lian, S.~Sokolov, E.~Y\"{u}ce, S.~Combri\'e, A.~De~Rossi, and A.~P. Mosk,
  \enquote{Dispersion of coupled mode-gap cavities,} Optics Letters
  \textbf{40}, 4488 (2015).

\bibitem{sokolov_measurement_2016}
S.~Sokolov, J.~Lian, S.~Combri{\'e}, A.~De~Rossi, and A.~P. Mosk,
  \enquote{Measurement of the linear thermo-optical coefficient of
  ${Ga}_{0.51}{In}_{0.49}{P}$ using photonic crystal nanocavities,}
  arXiv:1612.05544 [cond-mat, physics:physics]  (2016). ArXiv: 1612.05544.

\bibitem{tran2009photonic}
Q.~V. Tran, S.~Combri{\'e}, P.~Colman, and A.~De~Rossi, \enquote{Photonic
  crystal membrane waveguides with low insertion losses,} Applied Physics
  Letters \textbf{95}, 061105 (2009).

\bibitem{gottesman2004new}
Y.~Gottesman, E.~Rao, and D.~Rabus, \enquote{New methodology to evaluate the
  performance of ring resonators using optical low-coherence reflectometry,}
  Journal of lightwave technology \textbf{22}, 1566--1572 (2004).

\bibitem{sanogo2013phase}
Y.~Sanogo, A.-F. Obaton, C.~Delezoide, J.~Lautru, M.~Li{\`e}vre, J.~Dubard,
  I.~Ledoux-Rak, and C.~Nguyen, \enquote{Phase sensitive-optical low coherence
  interferometer: A new protocol to evaluate the performance of optical
  micro-resonators,} Journal of Lightwave Technology \textbf{31}, 111--117
  (2013).

\bibitem{parini2008time}
A.~Parini, P.~Hamel, A.~De~Rossi, S.~Combri{\'e}, Y.~Gottesman, R.~Gabet,
  A.~Talneau, Y.~Jaouen, G.~Vadala \emph{et~al.}, \enquote{Time-wavelength
  reflectance maps of photonic crystal waveguides: a new view on
  disorder-induced scattering,} Journal of Lightwave Technology \textbf{26},
  3794--3802 (2008).

\bibitem{combrie2007investigation}
S.~Combri{\'e}, N.-V.-Q. Tran, E.~Weidner, A.~De~Rossi, S.~Cassette, P.~Hamel,
  Y.~Jaou{\"e}n, R.~Gabet, and A.~Talneau, \enquote{Investigation of group
  delay, loss, and disorder in a photonic crystal waveguide by low-coherence
  reflectometry,} Applied physics letters \textbf{90}, 231104 (2007).

\bibitem{huang_optical_1991}
D.~Huang, E.~A. Swanson, C.~P. Lin, J.~S. Schuman, W.~G. Stinson, W.~Chang,
  M.~R. Hee, T.~Flotte, K.~Gregory, C.~A. Puliafito, and {others},
  \enquote{Optical coherence tomography,} Science (New York, NY) \textbf{254},
  1178 (1991).

\bibitem{gottesman2010time}
Y.~Gottesman, S.~Combri{\'e}, A.~DeRossi, A.~Talneau, P.~Hamel, A.~Parini,
  R.~Gabet, Y.~Jaouen, B.-E. Benkelfat, and E.~V. Rao, \enquote{Time-frequency
  analysis for an efficient detection and localization of side-coupled cavities
  in real photonic crystals,} Journal of Lightwave Technology \textbf{28},
  816--821 (2010).

\bibitem{sancho2012integrable}
J.~Sancho, J.~Bourderionnet, J.~Lloret, S.~Combri{\'e}, I.~Gasulla, S.~Xavier,
  S.~Sales, P.~Colman, G.~Lehoucq, D.~Dolfi \emph{et~al.}, \enquote{Integrable
  microwave filter based on a photonic crystal delay line,} Nature
  communications \textbf{3}, 1075 (2012).

\bibitem{colman2012blue}
P.~Colman, S.~Combri{\'e}, G.~Lehoucq, A.~De~Rossi, and S.~Trillo,
  \enquote{Blue self-frequency shift of slow solitons and radiation locking in
  a line-defect waveguide,} Physical review letters \textbf{109}, 093901
  (2012).

\bibitem{martin2017gainp}
A.~Martin, D.~Sanchez, S.~Combri{\'e}, A.~de~Rossi, and F.~Raineri,
  \enquote{{GaInP} on oxide nonlinear photonic crystal technology,} Optics
  Letters \textbf{42}, 599--602 (2017).

\bibitem{mann_theory_2015}
N.~Mann, A.~Javadi, P.~D. García, P.~Lodahl, and S.~Hughes, \enquote{Theory
  and experiments of disorder-induced resonance shifts and mode-edge broadening
  in deliberately disordered photonic crystal waveguides,} Physical Review A
  \textbf{92}, 023849 (2015).

\bibitem{thyrrestrup_statistical_2012}
H.~Thyrrestrup, S.~Smolka, L.~Sapienza, and P.~Lodahl, \enquote{Statistical
  {Theory} of a {Quantum} {Emitter} {Strongly} {Coupled} to
  {Anderson}-{Localized} {Modes},} Physical Review Letters \textbf{108} (2012).

\bibitem{topolancik_experimental_2007}
J.~Topolancik, B.~Ilic, and F.~Vollmer, \enquote{Experimental {Observation} of
  {Strong} {Photon} {Localization} in {Disordered} {Photonic} {Crystal}
  {Waveguides},} Physical Review Letters \textbf{99} (2007).

\bibitem{smolka_probing_2011}
S.~Smolka, H.~Thyrrestrup, L.~Sapienza, T.~B. Lehmann, K.~R. Rix, L.~S.
  Froufe-P\'erez, P.~D. García, and P.~Lodahl, \enquote{Probing the
  statistical properties of {Anderson} localization with quantum emitters,} New
  Journal of Physics \textbf{13}, 063044 (2011).

\bibitem{faggiani_lower_2016}
R.~Faggiani, A.~Baron, X.~Zang, L.~Lalouat, S.~A. Schulz, B.~O’Regan,
  K.~Vynck, B.~Cluzel, F.~de~Fornel, T.~F. Krauss, and P.~Lalanne,
  \enquote{Lower bound for the spatial extent of localized modes in
  photonic-crystal waveguides with small random imperfections,} Scientific
  Reports \textbf{6}, 27037 (2016).

\bibitem{xue_threshold_2016}
W.~Xue, Y.~Yu, L.~Ottaviano, Y.~Chen, E.~Semenova, K.~Yvind, and J.~Mork,
  \enquote{Threshold {Characteristics} of {Slow}-{Light} {Photonic} {Crystal}
  {Lasers},} Physical Review Letters \textbf{116} (2016).

\bibitem{minkov_statistics_2013}
M.~Minkov, U.~P. Dharanipathy, R.~Houdr\'e, and V.~Savona, \enquote{Statistics
  of the disorder-induced losses of high-{Q} photonic crystal cavities,} Optics
  Express \textbf{21}, 28233 (2013).

\bibitem{taguchi_statistical_2011}
Y.~Taguchi, Y.~Takahashi, Y.~Sato, T.~Asano, and S.~Noda, \enquote{Statistical
  studies of photonic heterostructure nanocavities with an average {Q} factor
  of three million,} Optics Express \textbf{19}, 11916 (2011).

\bibitem{van_driel_statistical_2007}
A.~F. van Driel, I.~S. Nikolaev, P.~Vergeer, P.~Lodahl, D.~Vanmaekelbergh, and
  W.~L. Vos, \enquote{Statistical analysis of time-resolved emission from
  ensembles of semiconductor quantum dots: {Interpretation} of exponential
  decay models,} Physical Review B \textbf{75} (2007).

\bibitem{sokolov_tuning_2016}
S.~Sokolov, J.~Lian, E.~Yüce, S.~Combri{\'e}, A.~De~Rossi, and A.~P. Mosk,
  \enquote{Tuning out disorder-induced localization in nanophotonic cavity
  arrays,} arXiv:1608.01257 [cond-mat, physics:physics]  (2016). ArXiv:
  1608.01257.

\bibitem{han2015high}
Z.~Han, G.~Moille, X.~Checoury, J.~Bourderionnet, P.~Boucaud, A.~De~Rossi, and
  S.~Combri{\'e}, \enquote{High-performance and power-efficient 2$\times$ 2
  optical switch on silicon-on-insulator,} Optics Express \textbf{23},
  24163--24170 (2015).

\end{thebibliography}

\end{document}